\documentclass[twoside, twocolumn, prb
 floats,
 amsmath,amssymb,
 aps,
]{revtex4-1}

\usepackage[english]{babel}
\usepackage[paperwidth=210mm,paperheight=297mm,centering,hmargin=2cm,vmargin=2.5cm]{geometry}
\usepackage{graphicx}
\usepackage{diagbox}
\usepackage[colorlinks=true, allcolors=blue]{hyperref}
\usepackage{changes}
\usetikzlibrary{patterns}

\makeatletter
\renewcommand*\env@matrix[1][c]{\hskip -\arraycolsep
  \let\@ifnextchar\new@ifnextchar
  \array{*\c@MaxMatrixCols #1}}
\makeatother

\begin{document}


\title{A minimal model of an artificial topological material realized in a two-terminal Josephson junction threaded by Aharonov-Casher fluxes}%

\author{Luka Medic}
 \email{luka.medic@ijs.si}
 \author{Anton Ram\v{s}ak}%
\author{Toma\v{z} Rejec}%
\affiliation{Jo\v{z}ef Stefan Institute, Jamova 39, SI-1000 Ljubljana, Slovenia}
\affiliation{Faculty of Mathematics and Physics, University of Ljubljana, Jadranska	19, SI-1000 Ljubljana, Slovenia}
 

\date{\today}

\begin{abstract}
    We investigate a minimal model of a two-terminal Josephson junction with conventional superconducting (SC) leads and a pair of interconnected quantum dots in the presence of two Aharonov-Casher (AC) fluxes. The Andreev bound state spectrum features Weyl nodes within a three-dimensional synthetic Brillouin zone defined in the space of these AC fluxes and the SC phase difference. The aim is to determine the location and topological charge of these nodes by probing the Berry curvature on closed surfaces that may enclose them. This is achieved by adiabatically varying the superconducting phase difference and AC fluxes along a path on these surfaces and measuring the associated currents. We define the kinematic curvature as the cross product of a tangent vector along the path and the vector of these currents. In the adiabatic regime, the path-averaged kinematic curvature provides a quantized response equal to the topological charge enclosed by the surface, provided the path uniformly and densely covers it.
\end{abstract}

\maketitle



\section{Introduction}

Advances in modern physics and technology have spurred great interest in the study of symmetry and topology in condensed matter physics \citep{chiu2016classification, hasan2010colloquium, qi2011topological}. Among these pursuits, significant attention has been drawn to Weyl semimetals (WSMs) \citep{hasan2017discovery, armitage2018weyl}, which host topologically protected Weyl nodes, leading to anomalous phenomena in these materials \citep{yang2011quantum, zyuzin2012topological, chernodub2014condensed}.

Treating independent superconducting (SC) phase differences as quasimomenta, an analog of WSMs can be realized in multi-terminal Josephson junctions \citep{riwar2016multi, eriksson2017topological}. In the subgap regime, the Andreev bound states (ABSs) exhibit Weyl singularities with conical dispersion. Importantly,  the authors of Ref. \citep{riwar2016multi} proposed a protocol and a measurable quantity capable of distinguishing between topological and trivial phases in such a system. Following their approach, one applies incommensurate voltages to two SC leads, causing the corresponding SC phase differences to traverse the entire two-dimensional synthetic Brillouin zone (BZ). The remaining independent SC phase differences serve as control parameters, enabling the transition of the system between distinct topological regimes, reflected in the change of transconductance, i.e. a time-averaged response in one SC lead due to voltage applied to another.

Subsequent research has explored systems where a SC control phase difference is replaced by a magnetic flux through the normal region  \citep{meyer2017nontrivial, xie2017topological, gavensky2019topological}. In contrast to Ref. \citep{riwar2016multi}, where at least four terminals are needed to realize Weyl topology, these works have shown that a three-terminal junction is sufficient in the presence of the magnetic flux.

In our recent investigation \citep{medic2024artificial}, we examined systems involving Aharonov-Casher (AC) effect \citep{aharonov1984topological, cimmino1989observation, mathur1992quantum, konig2006direct, bergsten2006experimental, liu2009control} where electrons with opposite spins acquire opposite AC phases \citep{bychkov1984oscillatory, manchon2015new, tomaszewski2018aharonov1, tomaszewski2018aharonov2}. We demonstrated that both the winding number and the Chern number can be identified within a two-terminal Josephson junction, and that, since the Weyl nodes are located only at SC phase difference $\phi=\pi$, the two topological invariants coincide.

In this paper, we analyze a toy model to explicitly demonstrate a topologically non-trivial regime within the class of systems introduced in Ref. \citep{medic2024artificial}. The system's three-dimensional synthetic BZ is in the space of two AC fluxes and the SC phase difference. Unlike SC phase differences, the AC flux in Rashba-gate-controlled rings \citep{citro2006pumping, wu2007analysis, grundler2000large, governale2003pumping} can only be varied over a limited range. Therefore, we consider driving protocols that vary the AC fluxes and the SC phase difference along a path on a small enough closed surface. To determine the enclosed topological charge, we introduce the concept of kinematic curvature, which is defined as the cross product of a tangent vector along the path and the vector of currents associated with AC fluxes and SC phase difference. We demonstrate the connection between the topological charge enclosed by the surface and the path-averaged kinematic curvature. Additionally, we propose a specific driving protocol in which the path is confined to the surface of a sphere and use it to compute the path-averaged kinematic curvature for the toy model.

The paper is structured as follows. In Sec \ref{sec:model}, we introduce a toy model incorporating two AC fluxes, present a phase diagram identifying gapless regimes that host Weyl nodes, and assign a topological charge to each node. In Sec \ref{sec:kinematicCurvature}, we introduce the kinematic curvature and demonstrate its connection to the topological charge. Finally, in Sec \ref{sec:drivingProtocolExample}, we provide a concrete example of a driving protocol in which the AC fluxes and SC phase difference evolve along a path covering the surface of a sphere.

\section{Model}
\label{sec:model}

\begin{figure}
	\includegraphics[width=0.5\textwidth]{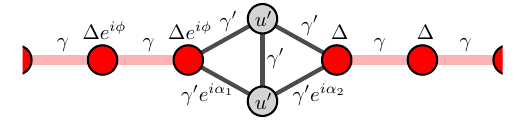}
	\caption{Superconducting leads (in red) attached to a pair of interconnected quantum dots (grey). The system is threaded by two AC fluxes, inducing phase factors $e^{i\alpha_1}$ and $e^{i\alpha_2}$.}
	\label{fig:systemSketch}
\end{figure}

Inspired by a model introduced in Ref. \citep{tomaszewski2018aharonov2}, we adopt a similar toy model featuring two SC leads coupled to two interconnected quantum dots (QDs), as illustrated in Fig. \ref{fig:systemSketch}. The Hamiltonian describing the QDs is
\begin{equation}
    H_{QD} = \begin{bmatrix}
         u' & -\gamma' \\
         -\gamma' & u' \\
    \end{bmatrix},
\end{equation}
where $u'$ represents the onsite potential on the QDs, and $\gamma'$ denotes the hopping between them. The hopping between the QDs, along with the tunneling from the SC leads to the QDs, defines two rings, each of which is threaded by an AC flux. As a result, an electron with spin $\uparrow$ moving in the anticlockwise direction along a ring acquires an AC phase, either $\alpha_1$ or $\alpha_2$, and the tunneling from the SC leads to the QDs can be expressed by the matrix
\begin{equation}
    H_T = -\gamma' \begin{bmatrix}
          1 & 1 \\
          e^{i \alpha_1} & e^{-i \alpha_2}\\
    \end{bmatrix}.
\end{equation}
In Ref. \citep{medic2024artificial}, we established that the topological charges for such a system are the same in both spin sectors. Therefore, we will only consider the spin sector with spin $\uparrow$ ($\downarrow$) electrons (holes).

The minimal model presented includes a single channel per lead, resulting in a $2 \times 2$ normal-state scattering matrix $S$. Neglecting its energy dependence \citep{medic2024artificial, beenakker1991universal}, it is given by
\begin{equation}
    S \equiv \begin{bmatrix}
        r & t' \\
        t & r' \\
    \end{bmatrix} = \left( \mathbb{I} - \frac{i}{\gamma} W \right)^{-1} \left( \mathbb{I} + \frac{i}{\gamma} W \right),
\end{equation}
where $r$ ($r'$) and $t$ ($t'$) denote the reflection and transmission amplitudes for a state incident on the scattering region from the left (right) lead, respectively, $\gamma$ is the hopping parameter in the leads, and $W = H_T^\dagger H_{QD}^{-1} H_T$. Closed-form expressions for the reflection and transmission amplitudes are provided in the Supplemental Materials (SM) S1 \citep{medic2024supplemental}.

\begin{figure}
	\includegraphics[width=0.5\textwidth]{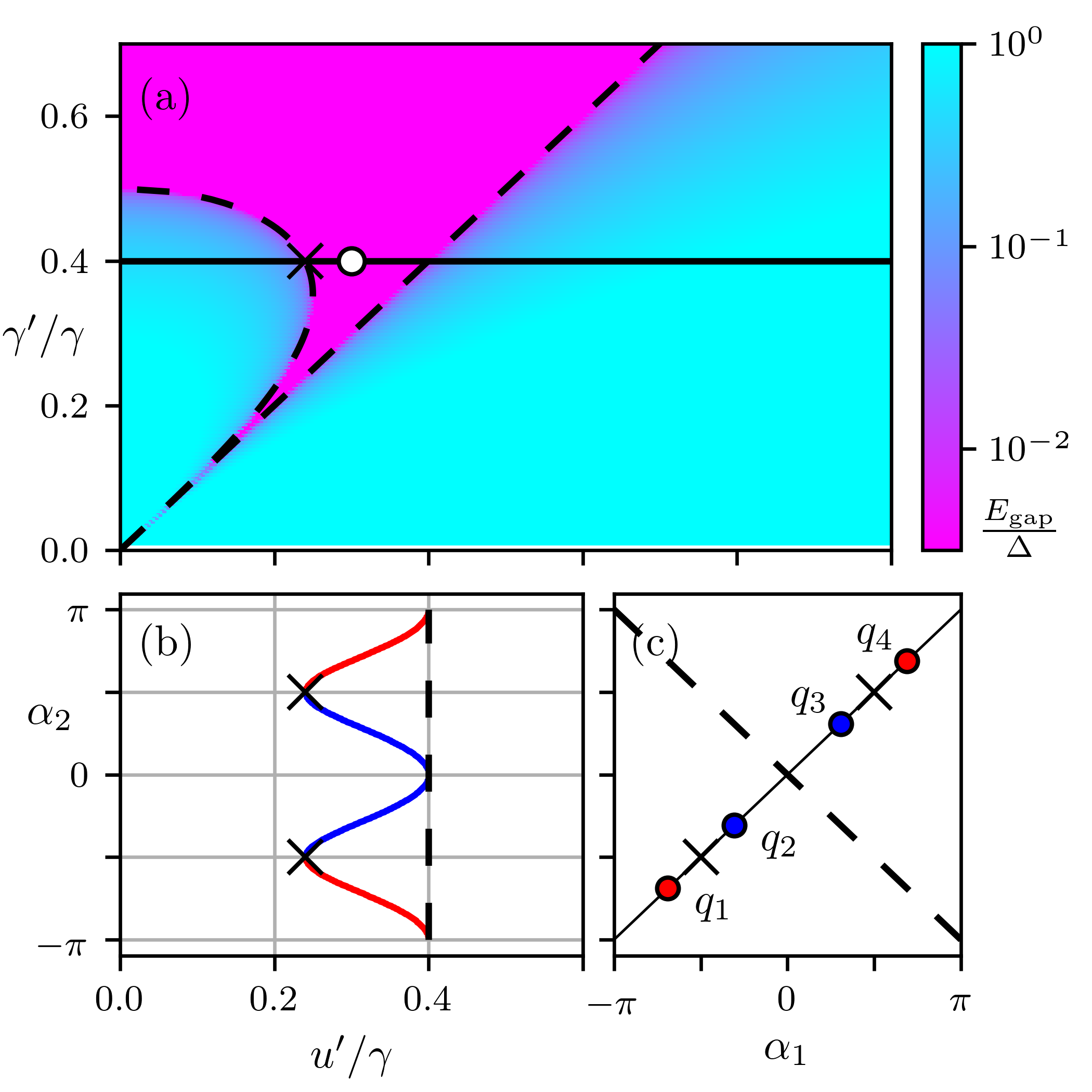}
	\caption{(a) ABS band gap in the parameter space $(u', \gamma')$. The dashed lines delineate the gapped and gapless regimes. (b) Positions of the Weyl nodes with $\gamma'/\gamma=0.4$ fixed and varying $u'$ [solid line in (a)]. Red (blue) lines represent paths of the Weyl nodes with positive (negative) topological charges. (c) Positions of the Weyl nodes in the synthetic BZ $(\alpha_1, \alpha_2)$ for $u'=0.3 \gamma$ and $\gamma'/\gamma=0.4 $. At $u' = \gamma'$, the ABS energy gap closes along the dashed line.}
	\label{fig:parameterSpace}
\end{figure}

The energy dispersion of Andreev bound states (ABS) for a single-channel two-terminal Josephson junction is given by $E_\pm = \pm \Delta\sqrt{1-T \sin^2(\phi/2)}$, where $\Delta$ is the SC gap, $\phi$ is the SC phase difference, and $T=|t|^2$ is the transmission eigenvalue \citep{beenakker1991universal}. Fig. \ref{fig:parameterSpace}(a) illustrates the phase diagram in the parameter space of $u'$ and $\gamma'$, showing the energy gap $E_{\textrm{gap}}$ between the two bands of ABSs in the synthetic BZ. The dashed lines in Fig. \ref{fig:parameterSpace}(a) delineate the gapped regime (in cyan) from the gapless one (purple); for details see SM S1 \citep{medic2024supplemental}. As we increase $u'$ while keeping $\gamma'$ fixed (e.g. $\gamma'=0.4 \gamma$), two pairs of Weyl nodes with opposite charges emerge at $\alpha_1=\alpha_2 = \pm \pi/2$ [indicated by crosses in Figs. \ref{fig:parameterSpace}(b) and (c)]. Further increasing the parameter $u'$ leads to the separation of the Weyl nodes in the space of AC fluxes, as depicted in Figs. \ref{fig:parameterSpace}(b) and \ref{fig:parameterSpace}(c). At $u' = \gamma'$, the Weyl nodes with the negative (positive) charge merge at $\alpha_{1,2} = 0$ ($\pi$). Simultaneously, the ABS band gap closes along $\alpha_1=-\alpha_2$ [dashed line in Fig. \ref{fig:parameterSpace}(c)], resulting in the annihilation of the charges, and the gap forms upon further increasing $u'$.

\begin{figure}
	\includegraphics[width=.5\textwidth]{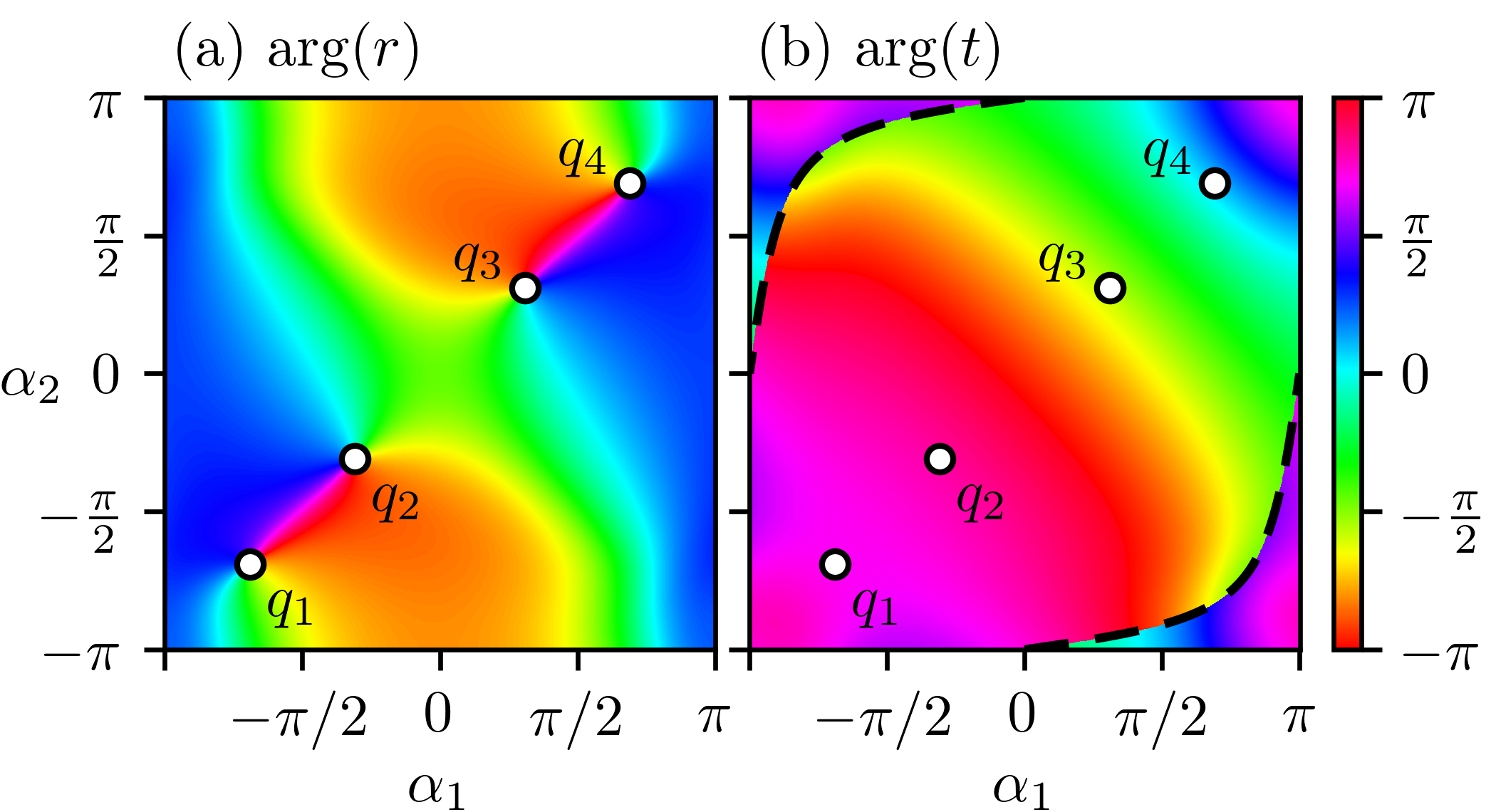}
	\caption{Phases of (a) the reflection amplitude $\textrm{arg}(r)$ and (b) the transmission amplitude $\textrm{arg}(t)$, plotted as functions of the AC fluxes $\alpha_1$ and $\alpha_2$, for $u' = 0.3\gamma$ and $\gamma' = 0.4\gamma$. Positions of Weyl nodes are indicated by $q_i$. Dashed lines represent a closed SC gap-edge-touching line.}
	\label{fig:scatteringMatrixPhases}
\end{figure}

Using the values of parameters corresponding to the white dot in Fig. \ref{fig:parameterSpace}(a), Fig. \ref{fig:scatteringMatrixPhases} illustrates the phases of reflection and transmission amplitudes, denoted as $\textrm{arg}(r)$ and $\textrm{arg}(t)$, respectively. The Weyl nodes are marked as $q_i$, and their topological charges can be visually inferred from Fig. \ref{fig:scatteringMatrixPhases}(a): encircling once a single Weyl node in the clockwise direction accumulates a phase of $\pm 2 \pi$, corresponding to a topological charge of $\pm 1$. To be more rigorous: following the methodology outlined in Ref. \citep{medic2024artificial}, the topological charge $q_{\scriptscriptstyle W}^{(i)}$ associated with the Weyl point $q_i$ at $\vec{x}_{\scriptscriptstyle W}^{(i)} = [\alpha_{\scriptscriptstyle W}^{(i)}, \alpha_{\scriptscriptstyle W}^{(i)}, \pi]^T$ can be determined from $r$ alone (see SM S1 \citep{medic2024supplemental} for detailed derivation):
\begin{align}
    q_{\scriptscriptstyle W}^{(i)} &= \mathrm{sgn}\left[\textrm{Im}\left(\frac{\partial r}{\partial \alpha_1}\cdot\frac{\partial r^*}{\partial \alpha_2}\right)\right]_{\alpha_{1,2}=\alpha_{\scriptscriptstyle W}^{(i)}} \nonumber \\
    &= -\mathrm{sgn}\left[\cos\left(\alpha_{\scriptscriptstyle W}^{(i)}\right)\right].
\end{align}

\section{Kinematic curvature}
\label{sec:kinematicCurvature}

We propose an observable, termed \textit{kinematic curvature}, and show that it can be used to determine the topological charge $q_{\scriptscriptstyle W}$. To develop this concept, we first define a vector of currents as $\vec{I} = \langle \psi | \nabla_{\vec{x}} H | \psi \rangle$, where $\vec{x} = [\alpha_1, \,\alpha_2, \,\phi]^T$ is a vector containing the AC fluxes and the SC phase difference, and the effective (electron) Hamiltonian acting in the ABS subspace is given by \citep{medic2024artificial}
\begin{equation}
    \label{eq:effectiveElectronHamiltonian}
    \!\! H \! = \! \Delta \! \begin{bmatrix}
        -(t' t'^\dagger)^{1/2} \cos\frac{\phi}{2} & r t^\dagger (t t^\dagger)^{-1/2} e^{i \phi/2} \\
        (t t^\dagger)^{-1/2} t r^\dagger e^{-i \phi/2} & (t t^\dagger)^{1/2} \cos\frac{\phi}{2} \\
    \end{bmatrix}\!.
\end{equation}
The first (second) component of $\vec{I}$ is proportional to the spin current in the left (right) ring, while the last component is proportional to the SC current. The wave function $|\psi\rangle$, initially prepared in the ground state, evolves according to the time-dependent Schrödinger equation:
\begin{equation}
    i\hbar |\dot{\psi}(\tau)\rangle = H(\tau) |\psi(\tau) \rangle,
\end{equation}
where $\tau$ denotes the time variable.

Assume that $\vec{x}(\tau)$ evolves along a path $\mathcal{P}$ that lies on a closed surface $\mathcal{S}$ within the synthetic BZ, which may potentially contain a Weyl node (or several of them). Let us define the kinematic curvature as
\begin{equation}
    \mathcal{\vec{F}} = \frac{ \dot{\vec{x}} }{ |\dot{\vec{x}}|^2 } \times \vec{I}.
\end{equation}
We will see that this quantity, which arises from the system's transport along a path, is closely related to the Berry curvature. For adiabatic driving, the currents can be expressed up to the first-order correction in velocity $\dot{\vec{x}}$ as \citep{riwar2016multi}
\begin{equation}
    \label{eq:vectorOfCurrrents_adiabaticApproximation}
    \vec{I} = \frac{1}{\hbar} \nabla_{\vec{x}} E_{-} - \dot{\vec{x}} \times \mathcal{\vec{B}}.
\end{equation}
The first term represents the adiabatic current, while the second term accounts for the first-order correction, with $\mathcal{\vec{B}} = -\textrm{Im} \left(\langle \nabla_{\vec{x}} \psi | \times | \nabla_{\vec{x}} \psi \rangle \right)$ denoting the Berry curvature. Then, the kinematic curvature can be expressed as
\begin{equation}
    \mathcal{\vec{F}} = \mathcal{\vec{B}} + \frac{1}{\hbar} \frac{ \dot{\vec{x}} }{ |\dot{\vec{x}}|^2 } \times \nabla_{\vec{x}} E_{-} - \frac{\dot{\vec{x}} \left(\dot{\vec{x}} \cdot \mathcal{\vec{B}}\right)}{|\dot{\vec{x}}|^2}.
\end{equation}
To eliminate the adiabatic term, one can assume a driving protocol in which the path $\mathcal{P}$ is traversed twice, once in each direction. Let $\hat{n}(\tau)$ denote the unit vector normal to the surface at the point $\vec{x}(\tau)$. Given that $\hat{n}$ and $\dot{\vec{x}}$ are orthogonal, we obtain $\langle \mathcal{\vec{F}} \cdot \hat{n} \rangle_{\mathcal{P}} = \langle \mathcal{\vec{B}} \cdot \hat{n} \rangle_{\mathcal{P}}$, where $\langle \, \rangle_{\mathcal{P}}$ denotes average along the path $\mathcal{P}$, which is traversed in both directions. [The average along $\mathcal{P}$ is defined as $\langle f \rangle_{\mathcal{P}} = \int_{\mathcal{P}} f(s) \, {\rm d}s \, \big/ \int_{\mathcal{P}} {\rm d}s$, where $s = \int_0^\tau |\dot{\vec{x}}(\tau')| \, {\rm d}\tau'$ is the arc length.] %

Assuming that the path evenly and densely covers the surface $\mathcal{S}$, allowing the line integral to be approximated by a surface integral, we derive that the \textit{path-averaged kinematic curvature}
\begin{equation}
    \label{eq:enclosedTopologicalCharge}
    \mathcal{Q} = \frac{|\mathcal{S}|}{2\pi}\big\langle \mathcal{\vec{F}} \cdot \hat{n} \big\rangle_\mathcal{P}
\end{equation}
is equal to the topological charge $\sum_{i} q_{\scriptscriptstyle W}^{(i)}$ enclosed inside $\mathcal{S}$. Here, $|\mathcal{S}|$ denotes the surface area of $\mathcal{S}$. Thus, by measuring the aforementioned currents, one can establish a connection between the system's topology and the (path-averaged) kinematic curvature.

Note that, to ensure adiabatic time evolution, the occupied ABSs must remain sufficiently separated from the unoccupied ones and the continuum states throughout the evolution. This requires that, along the protocol path, both Weyl nodes and SC gap-edge-touching singularities stay adequately distant from $\mathcal{S}$. Specifically, for the model introduced in the previous section, the Weyl nodes and an SC gap-edge-touching line (see SM S1 \citep{medic2024supplemental} for details) are indicated in Fig. \ref{fig:scatteringMatrixPhases}.

\section{Driving protocol example}
\label{sec:drivingProtocolExample}

As a specific example of a driving protocol, we consider the vector $\vec{x}$ confined to the surface of a sphere. The goal is to distinguish between spheres that enclose a Weyl point and those that do not.

\begin{figure}
	\includegraphics[width=.2\textwidth]{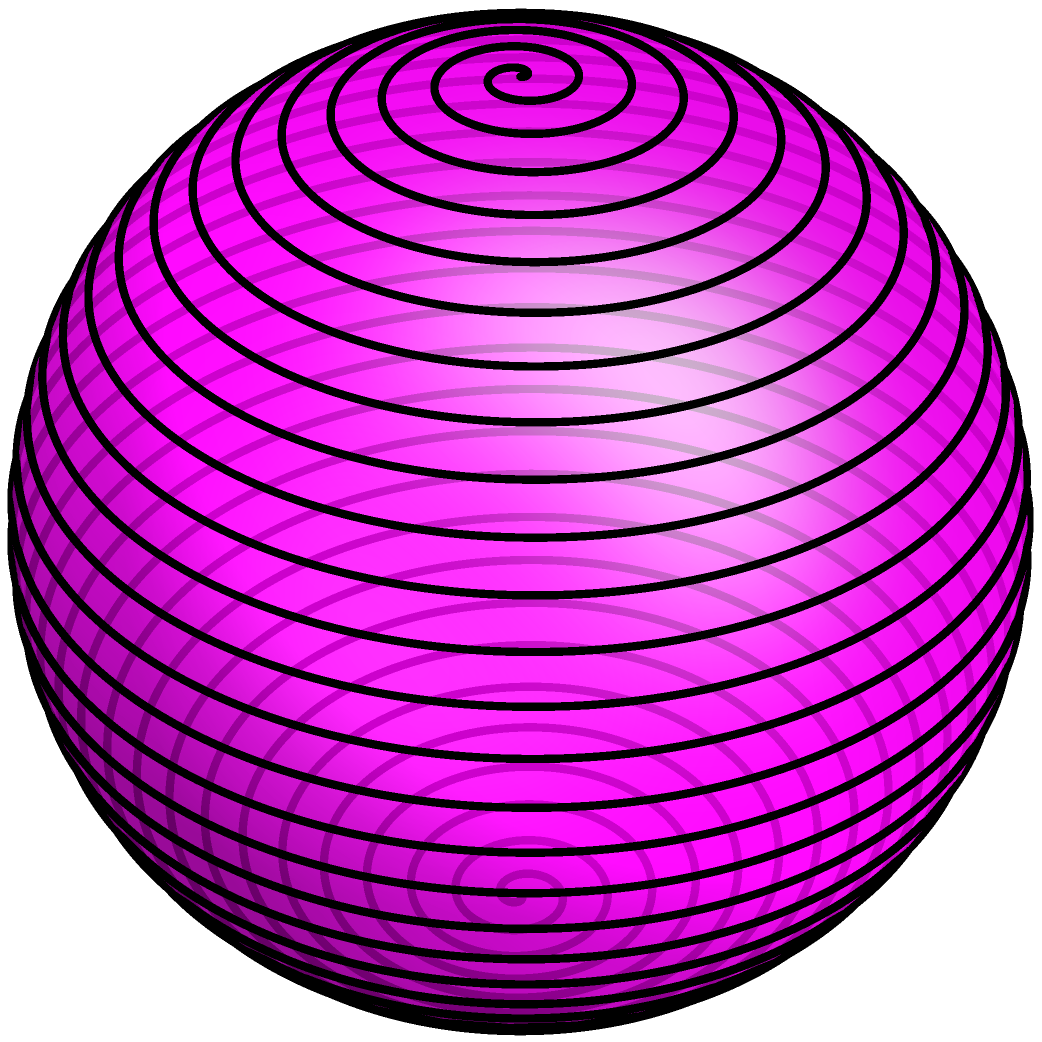}
	\caption{Depiction of the path on a sphere with $N=32$ revolutions around the polar axis.}
	\label{fig:plotSphereProtocol}
\end{figure}

A point on a sphere of radius $R$, centered at $\vec{x}_0$, is given by $\vec{x}(\theta, \phi) = \vec{x}_0 + R \, \hat{n}(\theta, \phi)$, where $\theta \in [0, \pi]$ and $\phi$ are the polar and azimuthal angles, respectively. To evenly traverse the sphere, we set $\phi(\theta) = 2\theta N$, where $N$ is the number of revolutions around the polar axis  (see Fig. \ref{fig:plotSphereProtocol}). For the polar angle, we choose $\theta(\tau) = \arccos\left( 1 - \frac{4\tau}{\tau_0} \right)$ for $\tau \in \left[0, \tau_0/2 \right]$, where $\tau_0$ is the total time to traverse the path in both directions. For $\tau \in \left[\tau_0/2, \tau_0\right]$, the path is retraced in the opposite direction. This choice of $\theta(\tau)$ ensures that the speed $ |\dot{\vec{x}}| $ remains approximately constant (with an average speed of $\langle|\dot{\vec{x}}|\rangle_\tau \approx 8NR/\tau_0$ for large $N$). Provided that $|\dot{\vec{x}}|$ remains small compared to the energy gap $\epsilon_{\textrm{gap}}$ along the path $\mathcal{P}$, the first-order correction in Eq. (\ref{eq:vectorOfCurrrents_adiabaticApproximation}) remains valid throughout the time evolution. [The only exception occurs near the poles, where $ |\dot{\vec{x}}| $ diverges. However, the contribution from this part of the path to the path-averaged kinematic curvature becomes negligible for large $N$. Moreover, more sophisticated protocols could easily be devised to mitigate the divergence in speed.]

\begin{figure}
	\includegraphics[width=.5\textwidth]{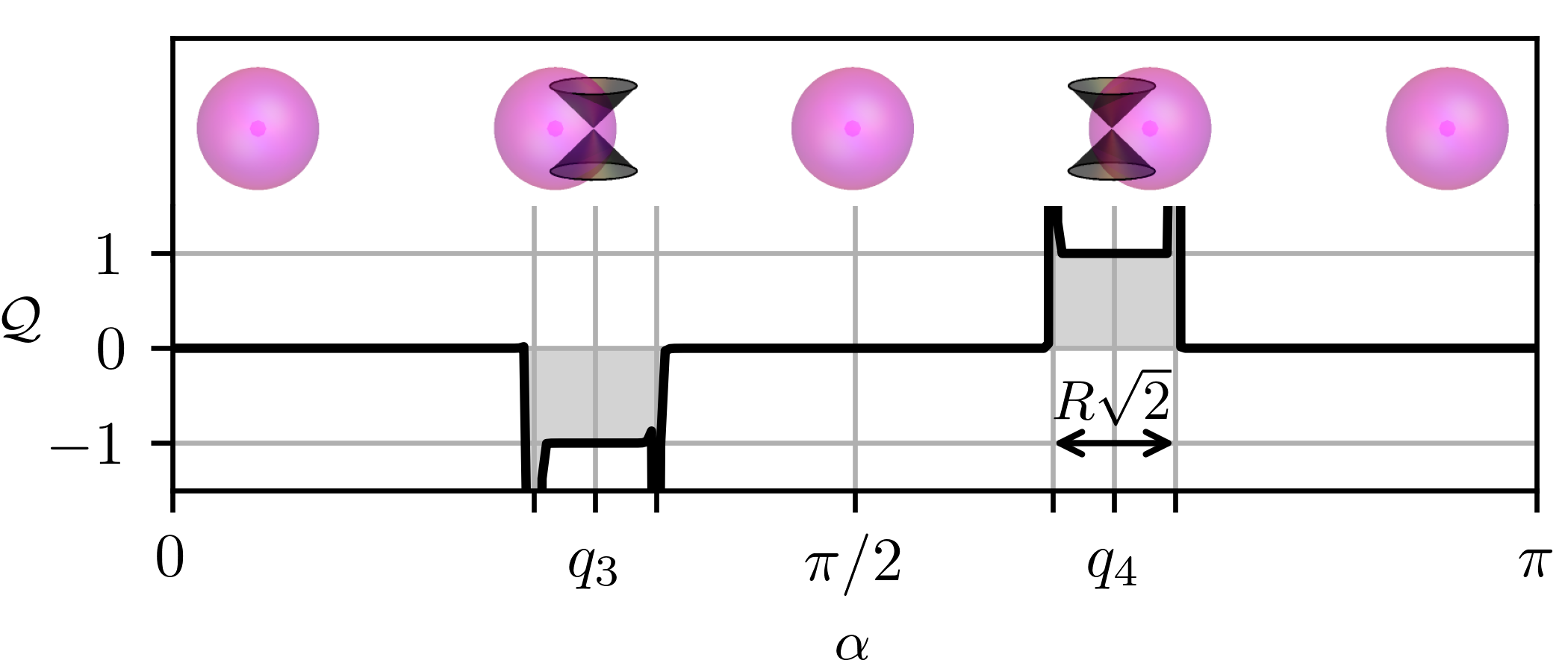}
	\caption{Quantization of $\mathcal{Q}$. The positions of the Weyl cones, and spheres (with radius $R$ and centers at $\vec{x}_0 = [\alpha, \alpha, \pi]^T$) on which the path $\mathcal{P}$ lies at a given $\alpha$ are shown above the plot.}
	\label{fig:topologicalCharge}
\end{figure}

We numerically verified Eq. (\ref{eq:enclosedTopologicalCharge}) by traversing spheres centered at $\vec{x}_0 = [\alpha, \alpha, \pi]^T$, where $\alpha$ specifies the center's position along the diagonal in the space of AC fluxes (cf. Fig. \ref{fig:scatteringMatrixPhases}). The quantization of $\mathcal{Q}$ is shown in Fig. \ref{fig:topologicalCharge}, which illustrates that paths on spheres enclosing Weyl nodes produce a non-zero quantized response, while those that do not enclose Weyl nodes yield a zero response. The width of the non-zero plateaus is $R \sqrt{2}$. We used the following parameters for the paths: $R=0.2$, $N=32$, and $\tau_0=2\cdot 10^6 \, \hbar / \Delta$. Near the edges of the plateaus, we observe spikes whose width decreases as $\tau_0$ increases (see SM S2 \citep{medic2024supplemental}). These spikes occur because certain sections of the path come close to a Weyl node, where $\epsilon_{\textrm{gap}}$ becomes small compared to $|\dot{\vec{x}}|$, causing the adiabatic approximation in Eq. (\ref{eq:vectorOfCurrrents_adiabaticApproximation}) to break down. The slight asymmetry between the spikes arises from deviations from conical dispersion as one moves away from the Weyl nodes and from the protocol's finite value of $N$.

In SM S2 \citep{medic2024supplemental}, an analysis with various values of $N$ and $\tau_0$ is presented, demonstrating the robustness of quantization even for small values of $N$, such as $N = 8$. The analysis also highlights the breakdown of quantization when the time per revolution around the polar axis, $\frac{\tau_0}{2N}$, approaches the characteristic timescale defined by $\epsilon_{\textrm{gap}}$.

In the Supplemental Materials, we demonstrate that our proposal extends to models with multi-channel leads. Furthermore, we present a comprehensive approach, working in the Nambu space to account for both electron and hole degrees of freedom, which yields results nearly identical to those in Fig. \ref{fig:topologicalCharge}, with only minor numerical deviations, especially near the spikes at the edges of the quantized plateaus. Finally, the results remain robust even when continuum states are included, which we verified numerically for systems with finite leads. Details of these computations can be found in SM S3-S5 \citep{medic2024supplemental}.

\section{Discussion}

In this paper, we presented an approach for experimentally determining the topological properties of a two-terminal Josephson junction with a normal region threaded by AC fluxes. Specifically, we investigated a model in which the normal region consists of two interconnected quantum dots. To measure the topological charges of Weyl nodes emerging in the ABS spectrum, we proposed a protocol involving a path in the space of SC phase difference and AC fluxes that covers a closed surface, potentially enclosing a topological charge. By traversing this path and measuring the superconducting and spin currents, we established a connection between the system's topology and the path-averaged kinematic curvature. Numerical simulations confirmed the effectiveness of the proposed protocol, demonstrating the quantization of $\mathcal{Q}$ with respect to the enclosed topological charge, where the paths enclosing Weyl nodes yield a non-zero quantized response.

For future investigation, it would be interesting to develop a protocol where the path is confined to the plane $\phi = \pi$ and to identify a measurable quantity directly linked to the winding number in the presence of chiral symmetry \citep{medic2024artificial}.

\section*{Acknowledgments}
The authors acknowledge support from the Slovenian Research and Innovation Agency (ARIS) under Contract No. P1-0044; AR was also supported by Grant J2-2514.

\bibliographystyle{apsrev4-1}
%

\end{document}



\title{Supplemental Materials\\[0.5cm]A minimal model of an artificial topological material realized in a two-terminal Josephson junction threaded by Aharonov-Casher fluxes}%

\author{Luka Medic}
 \email{luka.medic@ijs.si}
 \author{Anton Ram\v{s}ak}%
\author{Toma\v{z} Rejec}%
\affiliation{Jo\v{z}ef Stefan Institute, Jamova 39, SI-1000 Ljubljana, Slovenia}
\affiliation{Faculty of Mathematics and Physics, University of Ljubljana, Jadranska	19, SI-1000 Ljubljana, Slovenia}
 

\date{\today}

\maketitle



\onecolumngrid

\section{Analytical treatment of the model}

The closed-form expressions for the reflection and transmission amplitudes for an electron with spin $\uparrow$ incoming from the left lead are:
\begin{subequations}
    \begin{align}
    \label{eq:reflectionAmplitudeLeftLead}
    r(\alpha_1, \alpha_2) &=& \left( - 1 + \frac{2\gamma^2 \left((u')^2 - (\gamma')^2\right) - 4 i \gamma (\gamma')^2 \left(u' + \gamma' \cos(\alpha_2) \right) }{\gamma^2 \left((u')^2 - (\gamma')^2 \right) + 2 (\gamma')^4 \left(1 - \cos(\alpha_1 + \alpha_2)\right) + 2 i  \gamma (\gamma')^3 \left( \cos(\alpha_1)-\cos(\alpha_2)\right)} \right)^{-1} \\
    \label{eq:transmissionAmplitudeLeftLead}
    t(\alpha_1, \alpha_2) &=& \frac{2 i \gamma(\gamma')^2 \left[ u' \left(1+e^{i (\alpha_1+\alpha_2)}\right) + \gamma' \left(e^{i \alpha_1} + e^{i \alpha_2} \right)\right]}{\gamma^2 \left((u')^2 - (\gamma')^2\right) - 4 i \gamma (\gamma')^2 u' - 2 (\gamma')^4 \left( 1 - \cos(\alpha_1 + \alpha_2)\right) - 2 i \gamma(\gamma')^3 \left( \cos(\alpha_1) +\cos(\alpha_2)\right)}.
    \end{align}
\end{subequations}
Due to the left-right mirror symmetry of the model (see Fig. 1 of the main text), the reflection and transmission amplitudes for an incoming electron from the right lead are $r'(\alpha_1, \alpha_2) = r(-\alpha_2, -\alpha_1)$ and $t'(\alpha_1, \alpha_2) = t(-\alpha_2, -\alpha_1)$, respectively.

For a two-terminal junction, the energy dispersion is given by $E_\pm(\vec{x}) = \pm\sqrt{1-T(\alpha_1, \alpha_2) \sin^2(\phi/2)}$ \citep{beenakker1991universal}, where $T(\alpha_1, \alpha_2) = |t(\alpha_1, \alpha_2)|^2 = 1 - |r(\alpha_1, \alpha_2)|^2$. Therefore, ABS band touchings ($E_\pm=0$) occur when $\phi=\pi$ and $r=0$. Hence, we obtain the condition for the ABS band gap closure by setting the denominator in Eq. (\ref{eq:reflectionAmplitudeLeftLead}) to zero. Separating the real and imaginary parts, this leads to the conclusion that the system is gapless for $0 \leq \frac{\gamma^2\left((\gamma')^2 - (u')^2\right)}{4 (\gamma')^4} \leq 1$ at $\alpha_{\scriptscriptstyle W} = \alpha_1 = \alpha_2 = \frac{1}{2}\arccos\left(1 - \frac{\gamma^2\left((\gamma')^2 - (u')^2 \right)}{2 (\gamma')^4}\right)$.

The system exhibits SC gap edge touchings if and only if $\phi = 0$ or $t = 0$. The condition $t = 0$ is satisfied by setting the numerator of Eq. (\ref{eq:transmissionAmplitudeLeftLead}) to zero, leading to the following expression:
\begin{equation}
    e^{i \alpha_2} = - \frac{u' + \gamma' e^{i \alpha_1}}{\gamma' + u' \,e^{i \alpha_1}}.
\end{equation}
By satisfying this constraint separately for the real and imaginary parts, we obtain
\begin{equation}
    \alpha_2 = \pm \arccos\left(-\frac{2 u' \gamma' + \left((u')^2 + (\gamma')^2 \right) \cos(\alpha_1)}{(u')^2 + (\gamma')^2 + 2 u' \gamma' \cos(\alpha_1)}\right)
\end{equation}
with $+$ and $-$ signs for $\alpha_1 \in [-\pi, 0)$ and $\alpha_1 \in [0, \pi)$, respectively. Thus, instead of having SC gap-edge-touching nodes, the system possesses a closed SC gap-edge-touching line, as depicted in Fig. 3(b) of the main text (dashed).

For the derivatives of the reflection amplitude $r$, evaluated at $\alpha_{\scriptscriptstyle W} = \alpha_1 = \alpha_2$, which indicates the position of the Weyl node at $\vec{x}_{\scriptscriptstyle W} = [\alpha_{\scriptscriptstyle W}, \alpha_{\scriptscriptstyle W}, \pi]^T$, we compute
\begin{equation}
    \label{eq:reflectionDerivatives}
    \left.\frac{\partial r}{\partial \alpha_{1,2}}\right|_{\alpha_{1,2} = \alpha_{\scriptscriptstyle W}} = \frac{2 (\gamma')^3 \left(\gamma' \sin(2 \alpha_{\scriptscriptstyle W}) \mp i \gamma \sin(\alpha_{\scriptscriptstyle W}) \right)}{2 \gamma \left((u')^2 - (\gamma')^2 \right) - 4 i \gamma (\gamma')^2 \left(u' + \gamma' \cos(\alpha_{\scriptscriptstyle W}) \right)}
\end{equation}
where $-$ ($+$) sign is taken for the derivative with respect to $\alpha_1$ ($\alpha_2$).

Based on our previous work \citep{medic2024artificial}, we know that the topological charge $q_{\scriptscriptstyle W}$ can be computed as
\begin{equation}
    \label{eq:topologicalChargeFromZeta}
    q_{\scriptscriptstyle W} = \mathrm{sgn} \left[\mathrm{det}(M_2)\right] = \mathrm{sgn}\left[\mathrm{Im}\left(\frac{\partial \zeta}{\partial \alpha_1}\cdot\frac{\partial \zeta^*}{\partial \alpha_2}\right)\right]_{\vec{x}= \vec{x}_{\scriptscriptstyle W}},
\end{equation}
where
\begin{equation}
    M_2 = \begin{bmatrix}
    \partial_{\alpha_1} {\rm Re}(\zeta) & -\partial_{\alpha_1} {\rm Im}(\zeta) \\
    \partial_{\alpha_2} {\rm Re}(\zeta) & -\partial_{\alpha_2} {\rm Im}(\zeta) \end{bmatrix}_{\vec{x}= \vec{x}_{\scriptscriptstyle W}}, \qquad \zeta(\vec{x}) = \langle a_e^+ | H(\vec{x}) | a_e^- \rangle.
\end{equation}
Here, $H$ denotes the effective (electron) Hamiltonian within the ABS subspace [refer to Eq. (5) in the main text], and $|a_e^\pm \rangle$ represent the chiral states at the Weyl node with positive and negative chirality, respectively. For a $2 \times 2$ scattering matrix $S$, these states can be expressed as
\begin{equation}
    | a_e^+ \rangle = \begin{bmatrix}
    V_1(\vec{x}_{\scriptscriptstyle W}) \\
    0 \\
    \end{bmatrix}, \qquad | a_e^+ \rangle = \begin{bmatrix}
    0 \\
    V_2(\vec{x}_{\scriptscriptstyle W}) \\
    \end{bmatrix}.
\end{equation}
where $V_1 = r/|r| \equiv e^{i \theta_r}$ and $V_2 = -i \, t/|t| \equiv -i \, e^{i \theta_t}$ are complex numbers derived from the polar decomposition of $S$ \citep{martin1992wave, beenakker1997random}, which takes the form:
\begin{equation}
\label{eq:polarDecomposition}
    S \equiv \begin{bmatrix}
    r & t' \\
    t & r' \\
    \end{bmatrix} = \begin{bmatrix}
    V_1 & 0 \\
    0 & V_2 \\
    \end{bmatrix} \begin{bmatrix}
    -i |r| & |t| \\
    |t| & -i |r| \\
    \end{bmatrix} \begin{bmatrix}
    U_1^\dagger & 0 \\
    0 & U_2^\dagger \\
    \end{bmatrix}, \qquad U_1 = -i, \quad U_2 = r(t')^*/|r t| \equiv e^{i(\theta_{r} - \theta_{t'})}.
\end{equation}
Note that from the unitarity of $S$, it follows $r t^* + t' (r')^* = 0$. Since $r(\vec{x}_{\scriptscriptstyle W}) = 0$, the derivatives in Eq. (\ref{eq:topologicalChargeFromZeta}) reduce to
\begin{equation}
    \frac{\partial \zeta}{\partial \alpha_i}(\vec{x}_{\scriptscriptstyle W}) = \Delta e^{-i \theta_r(\vec{x}_{\scriptscriptstyle W})} \frac{\partial r}{\partial \alpha_i}(\vec{x}_{\scriptscriptstyle W}),
\end{equation}
which further simplifies the expression for the topological charge to
\begin{equation}
    q_{\scriptscriptstyle W} = \mathrm{sgn}\left[\mathrm{Im}\left(\frac{\partial r}{\partial \alpha_1}\cdot\frac{\partial r^*}{\partial \alpha_2}\right)\right]_{\alpha_{1,2}=\alpha_{\scriptscriptstyle W}}.
\end{equation}
Using Eq. (\ref{eq:reflectionDerivatives}), we can show that this evaluates to
\begin{equation}
    q_{\scriptscriptstyle W} = -\mathrm{sgn}\left[\sin(\alpha_{\scriptscriptstyle W}) \sin(2\alpha_{\scriptscriptstyle W})\right] = -\mathrm{sgn}\left[\cos(\alpha_{\scriptscriptstyle W})\right],
\end{equation}
which demonstrates the validity of Eq. (4) in the main text.

\vspace*{.5cm}
\section{Analysis of protocol parameters}

\begin{figure}
	\includegraphics[width=.9\textwidth]{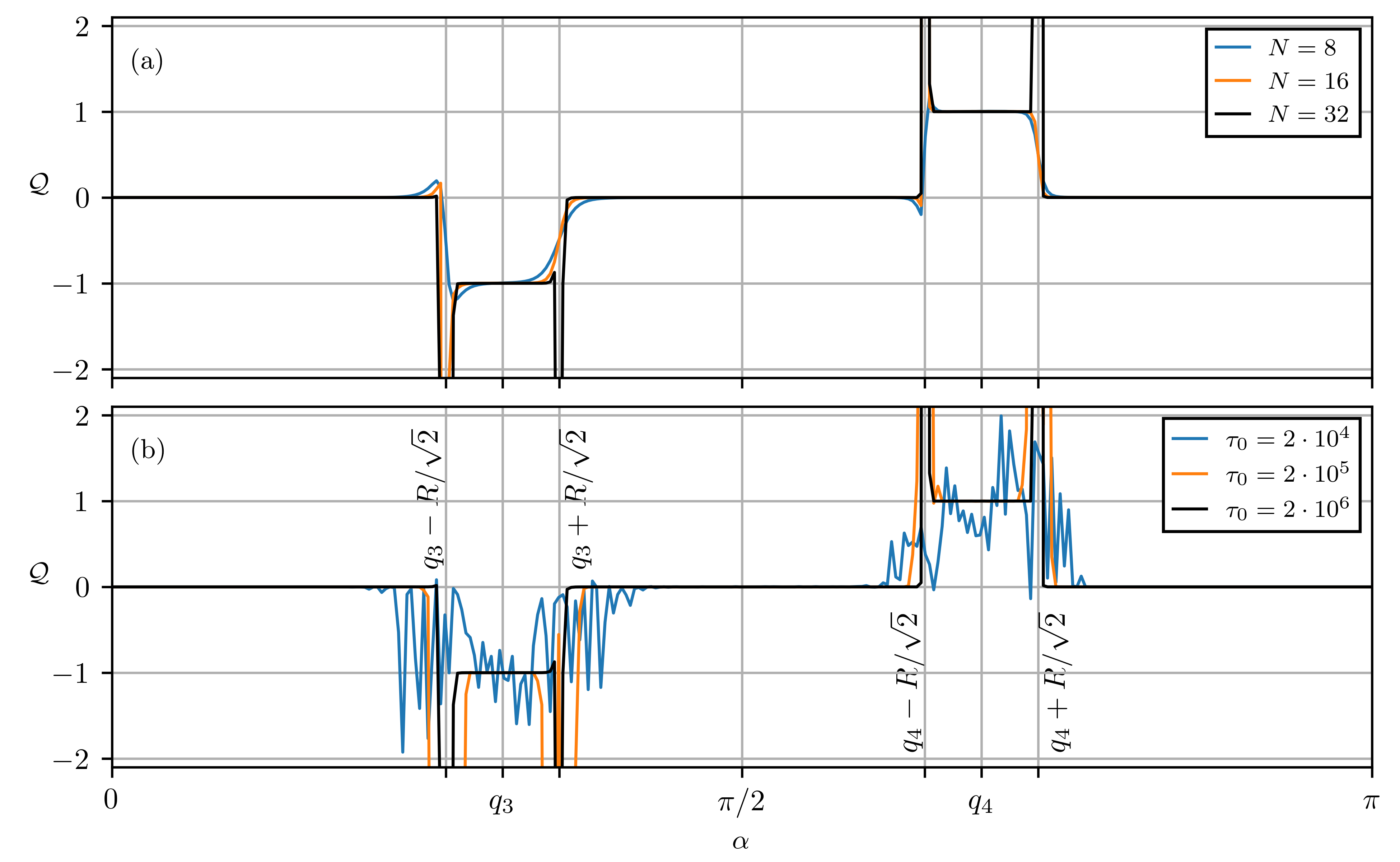}
	\caption{The path-averaged kinematic curvature $\mathcal{Q}$, with the center of the sphere at $\vec{x}_0 = [\alpha, \alpha, \pi]^T$. The parameters used are consistent with those in Fig. 5 of the main text. Panels (a) and (b) depict comparisons for distinct values of $N$ and $\tau_0$, respectively.}
	\label{fig:topologicalCharge_analysis}
\end{figure}

\begin{figure}
	\includegraphics[width=.9\textwidth]{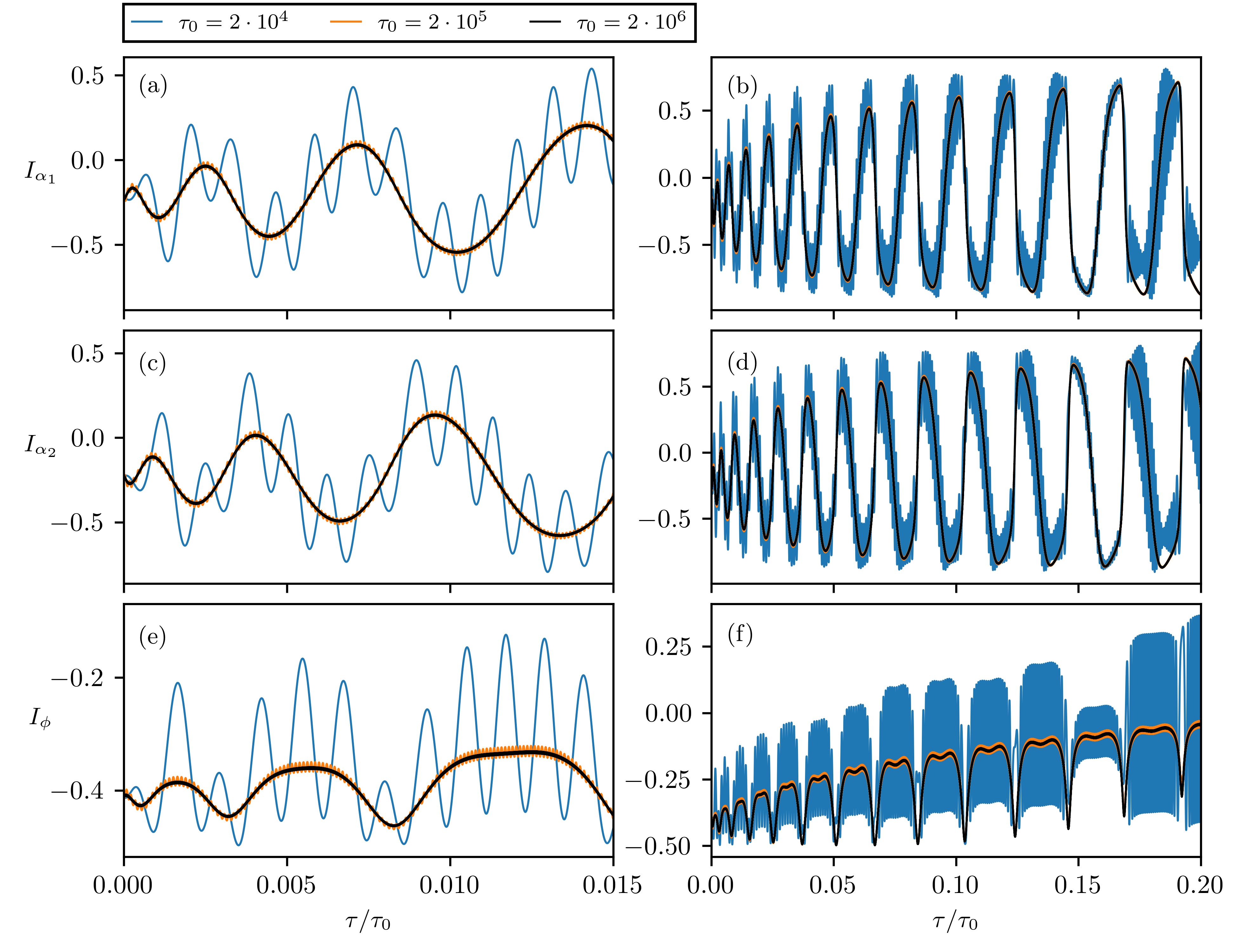}
	\caption{Currents for different driving periods $\tau_0$. The left and right panels display shorter and longer time windows, respectively. For $\tau_0 = 2\cdot 10^4$ (non-adiabatic regime), the two time scales, $\tau_{\textrm{gap}} \approx 30$ and $\tau_0 / (2N) \approx 300$, are comparable.}
	\label{fig:current_analysis}
\end{figure}

In the analysis of the protocol parameters, we vary the number of revolutions around the polar axis, $N$, and the duration of the driving protocol, $\tau_0$. The results are presented in Fig. \ref{fig:topologicalCharge_analysis}. Even for $N=8$, the driving protocol sufficiently samples typical values of the Berry curvature, enabling the recovery of the topological charge $q_{\scriptscriptstyle W}$. On the other hand, the correspondence between $\mathcal{Q}$ and the enclosed topological charge $q_{\scriptscriptstyle W}$ deteriorates when the assumption of adiabatic driving is not fulfilled for sufficiently short $\tau_0$. Specifically, as shown in Fig. \ref{fig:topologicalCharge_analysis}(b), this happens at $\tau_0 = 2 \cdot 10^4$ (in units of $\hbar/\Delta$). At this value, the time scale of one revolution around the polar axis, $\frac{\tau_0}{2 N} \approx 300$, becomes comparable to the time scale associated with the energy gap $\tau_{\textrm{gap}} = \frac{h}{\epsilon_{\textrm{gap}}} \approx 30$. The two time scales, with faster and slower oscillations, can be explicitly seen in the results for currents presented in Fig. \ref{fig:current_analysis}.

Additionally, in Fig. \ref{fig:topologicalCharge_analysis}, we observe that as $N$ decreases, the spikes at the edges of the quantized plateaus become less prominent or vanish entirely. This happens because lower values of $N$ reduce the likelihood of closely approaching the Weyl nodes, thereby better preserving the adiabatic assumption. Furthermore, the asymmetry in the shape of $\mathcal{Q}$ around each Weyl node $q_i$ becomes more pronounced. This increased asymmetry near the transitions results from the sparser path coverage of the sphere at smaller $N$.

\vspace*{.5cm}
\section{Extension to multi-channel leads}

\begin{figure}[h]
	\includegraphics[width=1.\textwidth]{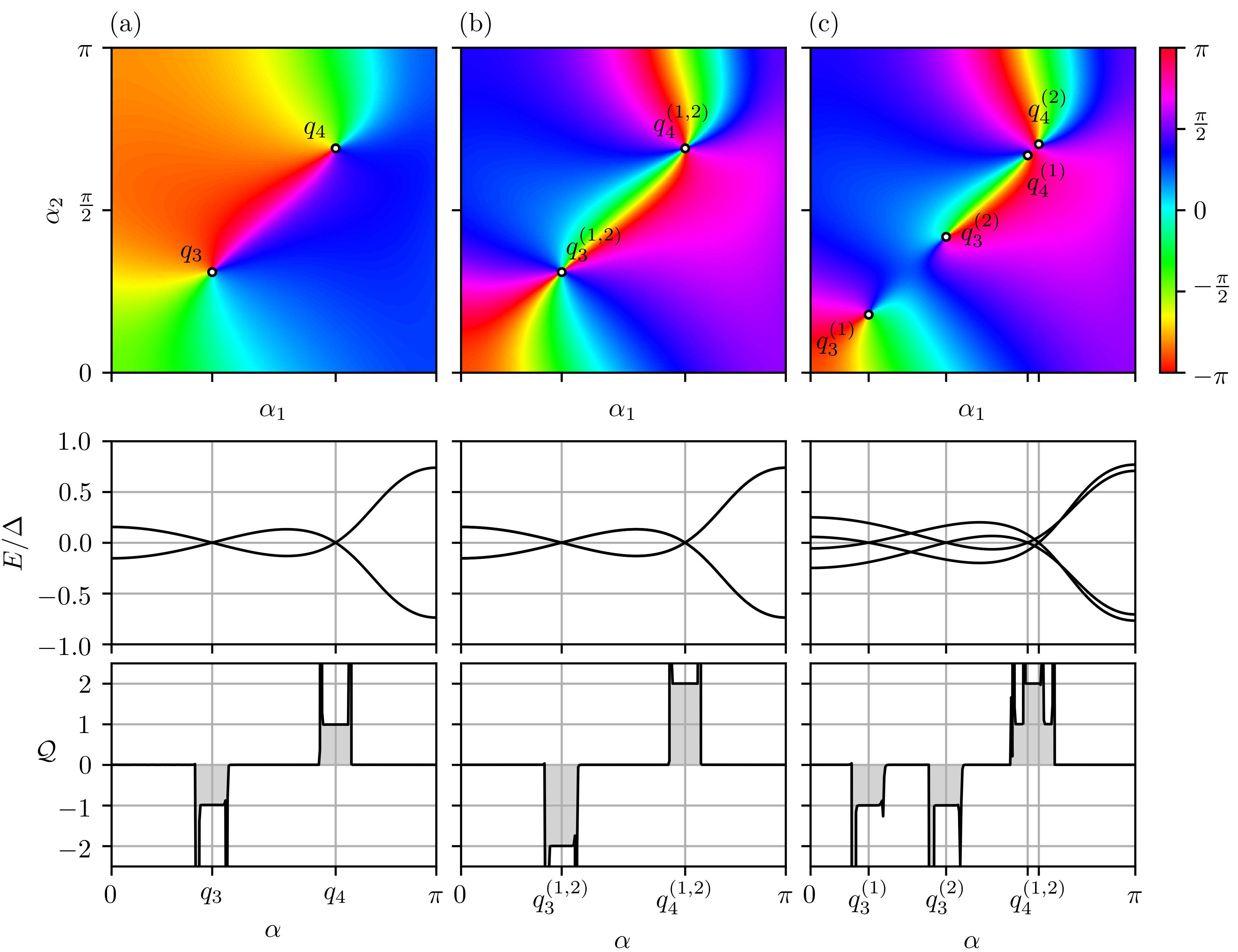}
	\caption{\textit{Top panels}: phases of the determinant of the reflection matrix at the left lead, ${\rm arg}(\det r)$, plotted in the plane of AC fluxes $\alpha_1$ and $\alpha_2$, with the positions of the Weyl nodes labeled by $q_i$.  \textit{Middle panels}: energy dispersions of the ABSs as a function of $\alpha = \alpha_1 = \alpha_2$. \textit{Bottom panels}: the path-averaged kinematic curvature $\mathcal{Q}$, illustrating its relationship to the Weyl node charges enclosed by a sphere centered at $(\alpha, \alpha, \pi)$. Plateaus correspond to quantized topological charges of $\pm 1$ or $\pm 2$, depending on whether the Weyl nodes are resolved individually or enclosed together.}
	\label{fig:scatteringMatrixPhasesAndCherns_multichannels}
\end{figure}

We extended our analysis to leads with multiple channels, specifically examining two coupled copies of the system described in the main text with inter-lead hopping, $\gamma_{\perp}$. In Fig. \ref{fig:scatteringMatrixPhasesAndCherns_multichannels}, we show results for (a) one copy, (b) two uncoupled copies, and (c) two coupled copies with $\gamma_{\perp} = 0.2 \gamma$. In contrast to the main text, where $r$ is a scalar, the top panels show ${\rm arg}\left({\rm det}\,r\right)$ \citep{medic2024artificial}, since, in general, $r$ is an $N_L \times N_L$ matrix, where $N_L$ represents the number of channels in the left lead. Notably, in Fig. \ref{fig:scatteringMatrixPhasesAndCherns_multichannels}(b), the complex phase of ${\rm det}\,r$ is doubled. When the inter-lead coupling is nonzero [Fig. \ref{fig:scatteringMatrixPhasesAndCherns_multichannels}(c)], nodes with topological charge $\pm 2$ split into pairs of Weyl nodes with charges $\pm 1$.

The middle panels display the ABS dispersions at $\phi = \pi$ along the diagonal $\alpha = \alpha_1 = \alpha_2$, while the bottom panels present the results of the driving protocol with the sphere's center positioned along this diagonal. For two uncoupled copies, the currents double, resulting in the measurement of topological charges $\pm 2$. For $\gamma_{\perp} = 0.2 \gamma$, the Weyl nodes $q_3^{(1,2)}$ are sufficiently far apart to be distinguishable by our protocol, whereas $q_4^{(1,2)}$ are separated by less than $R$ in the $(\alpha_1, \alpha_2)$ plane. With a sphere radius $R = 0.2$, both Weyl nodes can be enclosed simultaneously, leading to a plateau at $\mathcal{Q}=2$ in Fig. \ref{fig:scatteringMatrixPhasesAndCherns_multichannels}(c). Nonetheless, the Weyl nodes remain distinguishable, as indicated by narrow plateaus at $1$. The distinguishability of the nodes would improve in the adiabatic limit.

\section{Formulation of time evolution in the Nambu space}
\label{sec:nambuSpaceFormulation}

In Ref. \citep{medic2024artificial}, we introduced an alternative approach to computing topological invariants that accounts for the entire Nambu space, treating both the electron and hole components of ABSs rather than focusing solely on the electron part. Here, we extend this formalism to compute the time evolution of the wave function and the currents by utilizing the basis of instantaneous eigenfunctions and propagating the corresponding coefficients in time. Using these eigenfunctions as the basis is crucial, as the ABSs span only a submanifold of the full Nambu space.

We work in the instantaneous basis of the ABSs $| \varphi_n(\tau) \rangle$ of the time-dependent Bogoliubov-de Gennes (BdG) Hamiltonian $H(\tau)$:
\begin{equation}
    H(\tau) | \varphi_n(\tau) \rangle = E_n(\tau) | \varphi_n(\tau) \rangle.
\end{equation}
Solving the time-dependent equation
\begin{equation}
    i \hbar |\dot{\psi}(\tau) \rangle = H(\tau)|\psi(\tau) \rangle,
\end{equation}
where $|\psi(\tau) \rangle = \sum_n a_n(\tau) |\varphi_n(\tau)\rangle$ is written in the basis of instantaneous eigenfunctions, we obtain the system of differential equations governing the time evolution of the coefficients $a_n$:
\begin{equation}
    \dot{a}_m(\tau) = - i E_m(\tau) a_m(\tau) - \sum_n \langle \varphi_m(\tau) | \dot{\varphi}_n(\tau) \rangle a_n(\tau).
\end{equation}
To compute the nonadiabatic couplings $\langle \varphi_m(\tau) | \dot{\varphi}_n(\tau) \rangle$, a smooth structure gauge for the instantaneous eigenstates is required. This is obtained using polar decomposition \citep{medic2024artificial}. For the particular case under study, with only two ABSs ($m,n \in \{+,-\}$), the desired result is:
\begin{equation}
    \begin{bmatrix}
        \varphi_+ & \varphi_- \\
    \end{bmatrix} = \frac{1}{2} \begin{bmatrix}
        \varphi_\alpha & \varphi_\beta \\
    \end{bmatrix} \begin{bmatrix}
        e^{-i\phi/4} & 0 \\
        0 & e^{i\phi/4} \\
    \end{bmatrix} \begin{bmatrix}[r]
        1 & 1 \\
        1 & -1 \\
    \end{bmatrix} \begin{bmatrix}
        z/|z|& 0 \\
        0 & 1 \\
    \end{bmatrix} \begin{bmatrix}[r]
        1 & 1 \\
        1 & -1 \\
    \end{bmatrix}.
\end{equation}
Here, $z = |t|\cos(\phi/2) - i |r|$, and
\begin{equation}
    \label{eq:phi_alpha_beta}
    \begin{bmatrix}
        \varphi_\alpha & \varphi_\beta \\
    \end{bmatrix} = \frac{1}{\sqrt{2}} \begin{bmatrix}
        0 & r/|r| \\
        -it/|t| & 0 \\
        -i & 0 \\
        0 & -r (t')^*/|rt'|\\
    \end{bmatrix}
\end{equation}
where the scattering matrix elements $r$, $r'$, $t$, and $t'$ all change with the AC fluxes $\alpha_1$ and $\alpha_2$, which are time dependent. Expectation values for the currents are obtained from:
\begin{equation}
    \label{eq:currents_instantTimeEvolution}
    \vec{I}(\tau) = \langle \psi(\tau) | \nabla_{\vec{x}} H (\tau) | \psi(\tau) \rangle = \sum_n |a_n|^2 \nabla_{\vec{x}} E_n - \sum_{n \neq m} a_m^* a_n (E_m - E_n) \langle \varphi_m | \nabla_{\vec{x}}| \varphi_n \rangle.
\end{equation}
This approach yields results nearly identical to those in Fig. \ref{fig:topologicalCharge_analysis}, with only minor numerical deviations, especially near the transition points.

\section{Exact time evolution of the BdG Hamiltonian}

The effective electron Hamiltonian approach presented in the main text, along with its extension to the full Nambu space in Sec. \ref{sec:nambuSpaceFormulation}, introduces several approximations to the exact solution of our toy model. Namely, the Andreev approximation assumes that the SC band gap is small compared to the bandwidth ($\Delta \ll \gamma$), the energy dependence of the scattering matrix is neglected, and the contribution of continuum states to currents is disregarded.

\begin{figure}[h]
	\includegraphics[width=.55\textwidth]{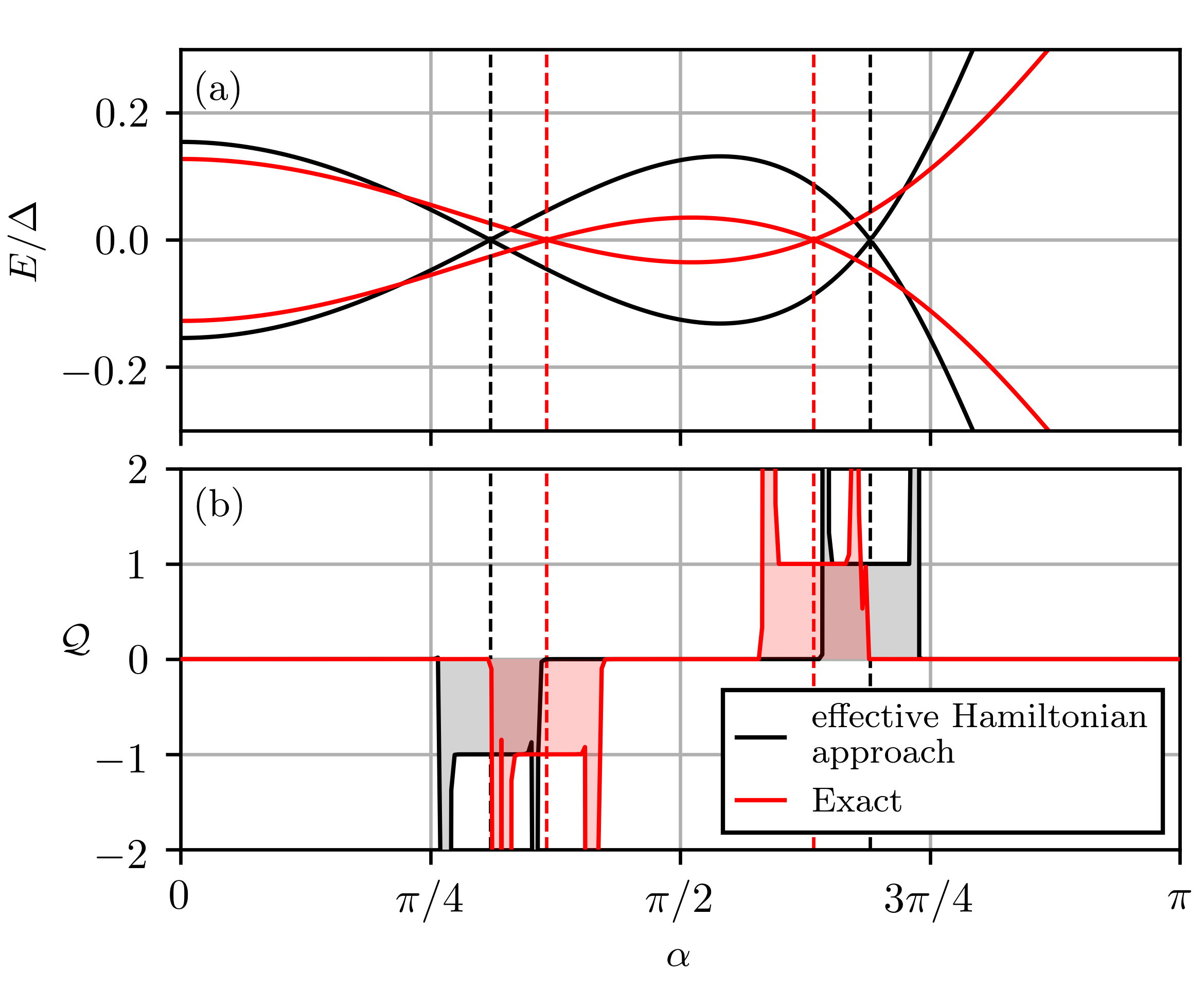}
	\caption{Comparison of the effective Hamiltonian approach presented in the main text with the exact solution: (a) energy dispersion and (b) quantization of $\mathcal{Q}$ as a function of $\alpha$, with spheres centered at $[\alpha, \alpha, \pi]^T$. Dashed lines indicate the positions of the Weyl nodes.}
	\label{fig:finiteLeads}
\end{figure}

To check that our conclusions do not rely on any of those approximations, we perform the exact time evolution of the BdG Hamiltonian for the toy model with a finite number of sites per SC lead. Starting with the ground state at $\tau=0$, we evolve the state in time and compute $Q$. The results for $\Delta=0.2\gamma$ and $L=50$ sites per SC lead are shown in Fig.~\ref{fig:finiteLeads}(b). Results are converged with respect to $L$. Note that the quantized plateaus in $Q$ are slightly shifted compared to those obtained using the effective Hamiltonian approach. This is due to the finite value of $\Delta$: the Andreev approximation is not strictly satisfied, resulting in a displacement of the Weyl nodes from their expected positions [see Fig.~\ref{fig:finiteLeads}(a)]. We performed the calculation both with and without accounting for continuum states (results including continuum states are not shown). The outcomes are indistinguishable, confirming that the quantization of $Q$ is exclusively due to ABS.

\bibliographystyle{apsrev4-1}
%